# Controlled Growth of a Line Defect in Graphene and Implications for Gate-Tunable Valley Filtering

J.-H. Chen[1,2,¶], G. Autès[3], N. Alem[1,2,δ], F. Gargiulo[3], A. Gautam[4], M. Linck[4,†], C. Kisielowski[4], O. V. Yazyev[3], S. G. Louie[1,2] and A. Zettl[1,2,*]

[1]Department of Physics, University of California at Berkeley, Berkeley, CA 94720, USA

[2]Materials Science Division, Lawrence Berkeley National Laboratory, Berkeley, CA 94720, USA

[3]Institute of Theoretical Physics, Ecole Polytechnique Fédérale de Lausanne (EPFL), CH-1015 Lausanne, Switzerland

[4]National Center for Electron Microscopy, Lawrence Berkeley National Laboratory, Berkeley, CA 94720, USA

*Correspondence to: azettl@berkeley.edu

[¶]Current Address: International Center for Quantum Materials, School of Physics, Peking University, Beijing, China 100871

[δ]Current Address: Department of Materials Science and Engineering, Pennsylvania State University, University Park, PA 16802, USA

[†]Current Address: CEOS-GmbH, Englerstr. 28, D-69126 Heidelberg, Germany

ABSTRACT: Atomically precise tailoring of graphene can enable unusual transport pathways and new nanometer-scale functional devices. Here we describe a recipe for the controlled production of highly regular "5-5-8" line defects in graphene by means of simultaneous electron irradiation and Joule heating by applied electric current. High-resolution transmission electron microscopy reveals individual steps of the growth process. Extending earlier theoretical work suggesting valley-discriminating capabilities of a graphene 5-5-8 line defect, we perform first-principles calculations of transport and find a strong energy dependence of valley polarization of the charge carriers across the defect. These findings inspire us to propose a compact electrostatically gated "valley valve" device, a critical component for valleytronics.

Atomically-precise modification of low-dimensional materials such as graphene is exceedingly challenging since existing experimental techniques rarely achieve atomic precision. Nevertheless, if successful, atomic manipulations could have dramatic impact on graphene's electrical, magnetic, optical, mechanical, chemical, and thermal properties[1-4], leading to novel functionalities that could be exploited in nanoscale devices. A recently emerging field is "valleytronics", a zero-magnetic-field analog to spintronics which exploits the quantum mechanical "valley" degree of freedom of charge carriers in graphene [2, 5-10].

At low energies the band structure of single-layer graphene is composed of two energetically degenerate valleys ("Dirac cones"), separated by ~30 $nm^{-1}$.[11] The intervalley coupling is quite weak in high quality graphene even at room temperature [12, 13], and hence this additional degree of freedom is a good quantum number. Valley polarization could be used for information processing much as the electron spin degree of freedom is used in spintronics, with the added benefit of temperature insensitivity. Generally, two approaches have been suggested for lifting the degeneracy and thus achieving graphene valley polarization: 1) application of external magnetic field, and 2) using local modifications of the crystalline lattice. The first approach requires either extreme operation conditions (e.g. very high magnetic fields and low temperatures [14, 15]) or bulky setups (e.g. optical pumping by circularly polarized light [16]) for the generation and detection of valley polarized currents. The second approach is more compatible with modern electronics fabrication techniques, but it requires a practical way for producing functional atomic-scale structures [2, 6, 7, 17] and a compact scheme for operating the device.

In this report, we show that the so-called "5-5-8" extended line defect can be produced in suspended graphene in a controlled way and in a pre-determined location without a catalyzing metal substrate. This defect structure, previously observed only in epitaxial graphene grown uncontrollably on a nickel substrate[18] is essentially a degenerate (zero misorientation angle) grain boundary consisting of alternating pentagon pairs and octagons (hence the 5-5-8 designation). Previous theoretical studies have suggested valley-discriminating transmission of charge carriers through the 5-5-8 line defect.[7] We further investigate the valley transport properties of this one-dimensional structural irregularity from first principles, and reveal a strong energy dependence of valley polarization of transmitted charge carriers. Based on this novel property, we propose the concept of an electrostatically operated graphene valley valve device.

The experiments are carried out in the ultra-high resolution aberration-corrected transmission electron microscope (TEM), the TEAM 0.5 at the National Center for Electron Microscopy at the Lawrence Berkeley National Lab. The acceleration voltage for the transmitted electrons is 80keV, below the ejection threshold of 86keV for fully bonded $sp^2$ carbon [19]. Single layer graphene is grown on a copper foil [20] and transferred to a silicon-based TEM sample chip (Protochips, Inc). Each TEM sample chip has a 500μm by 500μm silicon nitride window in the middle and pre-patterned large electrodes for *in-situ* electrical biasing. An array of holes 2μm in diameter is patterned in the silicon nitride window (before the transfer of graphene) and electron beam lithography is carried out to define the shape of the graphene flake as well as to make electrical contacts. After fabrication, the graphene devices are annealed in a hydrogen atmosphere for 2 hours at 350°C to remove PMMA residue [21]. The graphene sample is further cleaned inside the TEM by joule heating.

Figure 1 shows the experimental conditions for systematically producing 5-5-8 line defects. The key to successful controlled 5-5-8 defect formation is combining a directed electrical current with a free graphene edge. The free graphene edge is realized by means of a hole intentionally formed in the suspended graphene, and an electrical current density ($\sim 1.5\times 10^{10}$ Amp/m$^2$) is applied directly to the graphene sheet in a path traversing the hole. The current leads to local Joule heating of graphene to $T\sim 1300$K (see supplementary materials for the details of temperature estimation). Under such conditions, line defects nucleate from isolated pentagons formed at the edge of the hole (see insets i and ii in Figure 1), and then extend into the bulk of graphene by selective removal of carbon atoms from the pentagons. Higher atom mobility and less topological constraint near the graphene edge favors pentagon seeding, and thus the 5-5-8 line defects readily grow from the graphene edges near the hole generally in the direction of applied current.

Figure 2a shows a representative 5-5-8 line defect grown in graphene using the aforementioned hole/applied current method. The image represents the phase of a reconstructed TEM electron exit wave [22] made from a focal series of 80 images taken of the same area at different focus over about 60 seconds (see supplementary materials for details). Part of the hole in the graphene sheet is clearly seen in the upper right portion of the image (the defect always initiates at the graphene edge, but as it evolves, its "start" can recede from the edge). The inset to Fig. 2a is a magnified view of the rightmost portion of the line defect; the sequence of octagons alternated by pairs of pentagons is clearly visible. Fig. 2b is a schematic ball-and-stick representation of the same 5-5-8 defect, where the terminating carbon atoms and bonds are shown in red and the rest in orange. While the far right and left portions of the line defect in Fig. 2 strikingly display the expected 5-5-8 topology with clear vertically-stacked pentagon pairs, the center region is seemingly less distinct. As we discuss below, this "fuzziness" represents a structural resonance resulting from a topological frustration of the 5-5-8 line defect, and naturally occurs as the defect is growing. From Figures 2a and 2b it is apparent that the right (immobile) end of the defect is terminated by two heptagons and one pentagon (7-7-5 cluster), while the left (growth-leading) end is terminated by a pentagon-hexagon (5-6) pair. Calculations of defect formation energies show that the immobile 7-7-5 structure of line defect termination is energetically favorable over the 5-6 termination by ~1.5 eV (see supplementary materials for details). The extra 7-5 pair in the 7-7-5 structure serves as a dislocation which effectively relieves in-plane elastic strain associated with the line defect termination.

The growth process of the 5-5-8 linear defect is intriguing, and we examine it in some detail here. The metastable 5-6 pair generated at the growth-leading end is the key to the growth mechanism. Figures 3a-3c illustrate critical formation steps as determined by TEM. Each experimental image is constructed from an average of 12 single shot TEM images taken in rapid succession to reduce the background noise and to include all possible configurations of the defect. As the line defect grows by one octagon, one carbon atom (marked by a blue dot in the illustrations) is ejected, and a new bond is formed between its nearest neighbors (marked by yellow dots). This process also creates a new 5-6 termination pair, which serves as a seed for continued growth. Since an isolated pentagon cannot be sustained in otherwise ideal graphene [23], an extended structural irregularity such as the edge is required for the initial generation of pentagon-based seeds. Indeed, every 5-5-8 line defect we have created (more than 10) grows starting at the hole edges in graphene. As the line defect grows with a leading 5-6 pair, the other immobile end usually reconstructs into a very stable 7-7-5 cluster separated slightly from the graphene hole edge, as exemplified by Fig. 2a.

Figures 3a-3c also reveal the topologically stable and frustrated states during growth. The stable structure of the 5-5-8 line defect consists of alternating octagons and vertically stacked pentagon-pairs while the termination structure determines the boundary configuration of the line defect to be a 5-5 pair or an octagon. In the case of leading 5-6 pair and trailing 7-7-5 cluster terminations, a stable line defect requires an even total number of octagons and pentagon-pairs (a pair of pentagons counts as one unit, same as one octagon). Because the defect structure grows by the removal of one carbon atom at a time at the 5-6 termination, the line defect exists in two structural forms. The difference between the two is due to the fact that the structure of line defect is governed by the dimerization of carbon atoms along the defect. In case there is an even number of such carbon atoms (Figs. 3a and 3c) the structure of the defect is defined by a single minimum on the potential energy surface. However, a number of nearly degenerate configurations correspond to line defects characterized by an odd number of carbon atoms located along the line (Fig. 3b). Transitions between these configurations are likely to occur at a time scale much shorted the TEM image acquisition time. Thus, the observed structures appear "smeared out" without revealing any clear dimerization pattern as in Fig. 3b as well as in the central region of the defect shown in Figs. 2a,b. During the growth process, the observed defect structures alternate between the stable and frustrated states as the total number of carbon atoms along the defect alternates between even and odd numbers.

The controlled growth dynamics demonstrated in our experiment can in principle be exploited to fabricate arrays of 5-5-8 line defects in graphene for generating and detecting valley polarization. In order to better understand the electronic transport and valley filtering properties of the 5-5-8 line defect, we perform first-principles simulations of the electronic transport (see supplementary materials for details). Based on symmetry considerations, it has been previously argued that a 5-5-8 defect can act as a valley filter with valley selectivity depending on the angle of incidence $\theta$ of the charge carriers [7]. Fig. 4a shows the calculated band structure of graphene with a 5-5-8 line defect. We note the presence of several bands crossing the Dirac cone feature of the projected 2D band structure of graphene (shaded area in Fig. 4a). These bands correspond to electronic states localized at the line defect. Namely, there are two localized-state bands in the vicinity of the Dirac point ($E = 0$ eV) and one at significantly higher energies (0.5 eV < $E$ < 1 eV). The calculated transmission probabilities (Fig. 4b) show significant suppression of conductance at the Dirac point due to resonant backscattering of charge carriers by the states localized at the line defect [24]. One can expect that the valley filtering properties will eventually be dominated by this suppression of the charge-carrier transmission rather than by symmetry-based consideration. Figs. 4c and 4d show the values of angular-dependent valley polarization

$$P_\tau(\theta,E) = \frac{T_{\tau=+1}(\theta,E) - T_{\tau=-1}(\theta,E)}{T_{\tau=+1}(\theta,E) + T_{\tau=-1}(\theta,E)} \qquad (1)$$

calculated for electron and hole charge carriers, respectively. In this expression $T_{\tau=+1}(\theta,E)$ and $T_{\tau=-1}(\theta,E)$ are respective transmission probabilities of charge carriers belonging to the two valleys ($\tau=+1$ and $\tau=-1$) at incident angle $\theta$ and energy $E$. Indeed, we find that the dependences of $P_\tau(\theta,E)$ on $\theta$ are practically opposite to the ones predicted by the symmetry considerations (dashed lines in Figs. 4c and 4d) for the energies $|E| < 0.2$ eV. However, for the high-energy charge carriers not affected by the resonances, the behavior $P_\tau(\theta,E)$ predicted from symmetry arguments of Ref. 7 is mostly restored. Importantly, the revealed energy dependence opens another possibility for controlling the valley polarization of charge carriers in graphene with strong implications for graphene-based valleytronics.

Below we demonstrate the value of the energy dependence of valley transport properties by describing a concept of simple valleytronic device functioning in the ballistic regime. Figure 5 shows schematic illustration of electrically operated valley valve that is an analog of a standard spintronic device − the lateral spin valve [25, 26]. Unlike its spintronic counterpart, such a valley valve does not need magnetic leads and is simpler to operate. For example, by changing the local Fermi level from $E = 0.1$ eV to $E = 0.8$ eV relative to the charge-neutrality point of graphene, the valley polarization of initially unpolarized electron charge carriers incident at 30 degrees with respect to the normal direction of the first line defect can be switched from $P_\tau = -0.44$ (i.e. valley $\tau = -1$ polarized) to $P_\tau = +0.68$ (i.e. strongly valley $\tau = +1$ polarized). The second line defect serves as a valley-polarization detector, resulting in either a high resistance state (valley valve closed) or a low resistance state (valley valve opened).

In summary, we describe a method for the atomically precise engineering of 5-5-8 line defect in graphene which has been predicted to exhibit valley-discriminating transport properties. High-resolution transmission electron microscopy reveals individual steps of the defect formation process. We further employ first-principles calculations to demonstrate the energy dependence of its valley transport properties, which can be exploited in electrically switchable valleytronic devices such as valley filers and valves.

**Acknowledgements**

This work was supported in part by the Director, Office of Energy Research, Office of Basic Energy Sciences, Materials Sciences and Engineering Division, of the U.S. Department of Energy under Contract No. DE-AC02-05CH11231, under the sp2-bonded Materials Program, which provided for design of the experiment and analysis of TEM data; the Office of Naval Research for implementation of the biasing stage, and the National Science foundation under the Center of Integrated Nanomechanical Systems for postdoctoral support (N.A.) for sample preparation. TEAM0.5 microscope time was provided by the National Center for Electron Microscopy supported by the U.S. Department of Energy. G.A., F.G. and O.V.Y. acknowledge financial support of the Swiss National Science Foundation (grant No. PP002P_133552). First-principles computations have been performed at the Swiss National Supercomputing Centre (CSCS) under project s443. J.-H.C acknowledges helpful discussion with Qin Zhou on finite element simulation.

**References**

[1] Y.-W. Son, M. L. Cohen, and S. G. Louie, Phys. Rev. Lett. **97**, 216803 (2006).
[2] A. Rycerz, J. Tworzydlo, and C. W. J. Beenakker, Nat. Phys. **3**, 172 (2007).

3   A. V. Krasheninnikov, P. O. Lehtinen, A. S. Foster, P. Pyykkö, and R. M. Nieminen, Phys. Rev. Lett. **102**, 126807 (2009).
4   F. Banhart, J. Kotakoski, and A. V. Krasheninnikov, ACS Nano **5**, 26 (2011).
5   O. Gunawan, Y. P. Shkolnikov, K. Vakili, T. Gokmen, E. P. De Poortere, and M. Shayegan, Phys. Rev. Lett. **97**, 186404 (2006).
6   A. R. Akhmerov and C. W. J. Beenakker, Phys. Rev. Lett. **98**, 157003 (2007).
7   D. Gunlycke and C. T. White, Phys. Rev. Lett. **106**, 136806 (2011).
8   D. Xiao, G.-B. Liu, W. Feng, X. Xu, and W. Yao, Phys. Rev. Lett. **108**, 196802 (2012).
9   K. F. Mak, K. He, J. Shan, and T. F. Heinz, Nat. Nano. **7**, 494 (2012).
10  T. Cao, et al., Nat. Commun. **3**, 887 (2012).
11  P. R. Wallace, Phys. Rev. **71**, 622 (1947).
12  A. C. Ferrari, et al., Phys. Rev. Lett. **97**, 187401 (2006).
13  J.-H. Chen, W. G. Cullen, C. Jang, M. S. Fuhrer, and E. D. Williams, Physical Review Letters **102**, 236805 (2009).
14  Y. Zhang, et al., Phys. Rev. Lett. **96**, 136806 (2006).
15  Z. Zhu, A. Collaudin, B. Fauque, W. Kang, and K. Behnia, Nat. Phys. **8**, 89 (2011).
16  W. Yao, D. Xiao, and Q. Niu, Phys. Rev. B **77**, 235406 (2008).
17  D. Gunlycke, S. Vasudevan, and C. T. White, Nano Letters **13**, 259 (2013).
18  J. Lahiri, Y. Lin, P. Bozkurt, I. I. Oleynik, and M. Batzill, Nat. Nanotechnol. **5**, 326 (2010).
19  Ç. Ö. Girit, et al., Science **323**, 1705 (2009).
20  X. Li, et al., Science **324**, 1312 (2009).
21  M. Ishigami, J. H. Chen, W. G. Cullen, M. S. Fuhrer, and E. D. Williams, Nano Lett. **7**, 1643 (2007).
22  A. Thust, W. M. J. Coene, M. Op de Beeck, and D. Van Dyck, Ultramicroscopy **64**, 211 (1996).
23  O. V. Yazyev and S. G. Louie, Phys. Rev. B **81**, 195420 (2010).
24  H. J. Choi, J. Ihm, S. G. Louie, and M. L. Cohen, Phys. Rev. Lett. **84**, 2917 (2000).
25  R. Meservey, P. M. Tedrow, and P. Fulde, Physical Review Letters **25**, 1270 (1970).
26  P. M. Tedrow and R. Meservey, Physical Review Letters **26**, 192 (1971).

**Figures**

**Fig. 1 Growth of a 5-5-8 line defect from the edges of graphene under large electrical bias.**

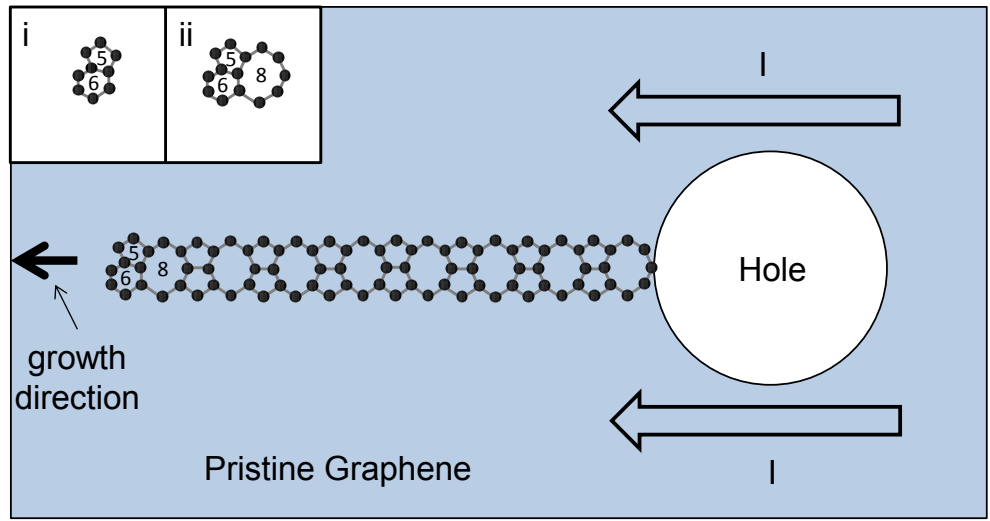

**Fig. 1 | Growth of a 5-5-8 line defect from the edges of graphene under large electrical bias.** Schematic drawing of the growth process of a 5-5-8 line defect in graphene. Graphene edges and a large electrical current are found to be the necessary condition for the 5-5-8 line defect to grow. Insets i and ii show the proposed initial stages of the line defect grown from a 5-6 pair at the edge of graphene (see Fig. 3 for experimental data and detailed growth steps).

**Fig. 2 Reconstructed exit-wave phase image of a 5-5-8 line defect in graphene**

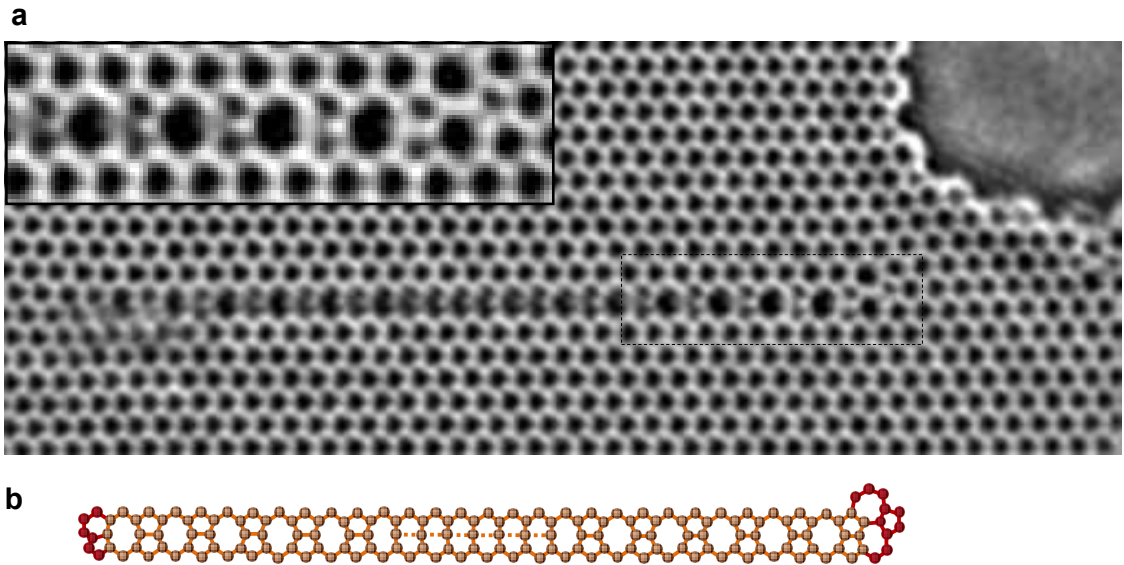

**Fig. 2 | Reconstructed exit-wave phase image of a 5-5-8 line defect in graphene. a.** The phase of the electron exit-wave from a 5-5-8 line defect in graphene was reconstructed using a focal series of 80 TEM images taken at the defect. **b**. Illustration of the atomic structure of the line defect (orange dots) and its termination (red dots).

**Fig. 3 Self-catalyzed growth of the 5-5-8 line defect**

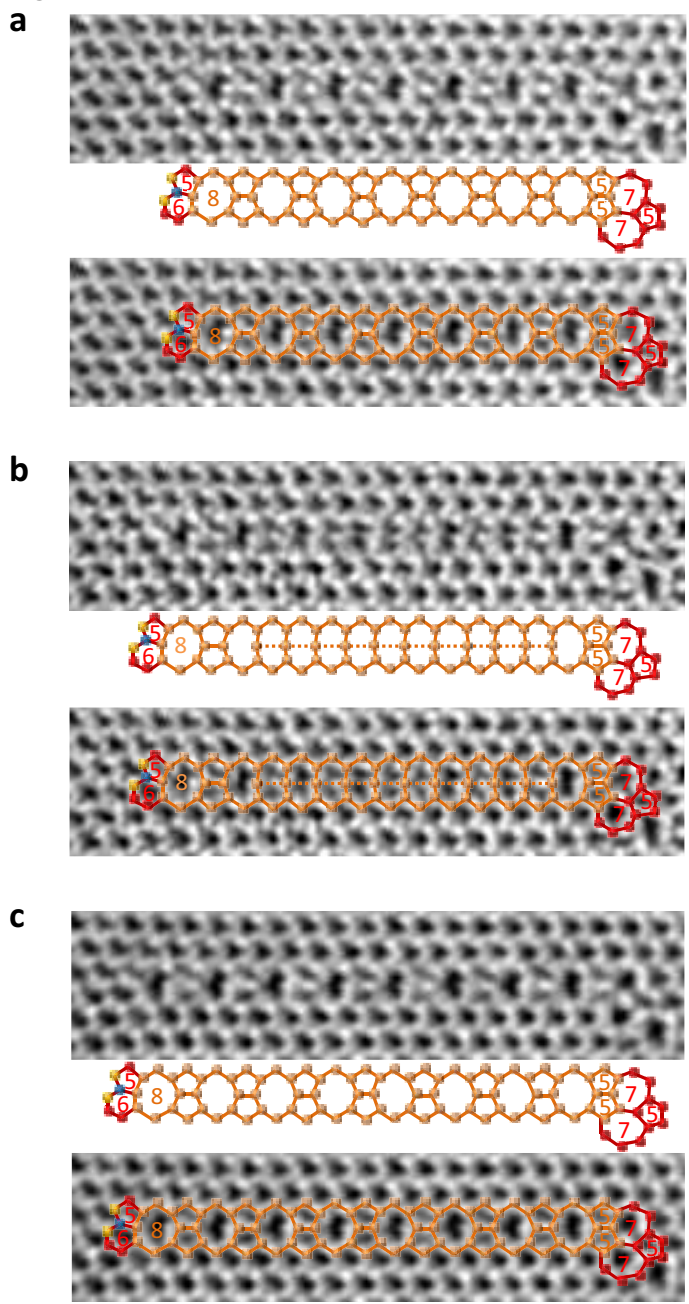

**Fig. 3 | Self-catalyzing growth of the 5-5-8 line defect. a-c.** Time-series of images capturing the growth of the line defect. Atomic structure models and their overlays onto the experimental data are also shown. The line defect grows by one octagon each time by the ejection of one carbon atom (marked by the blue dots in the illustrations) from the 5-6 termination and the formation of a new bond between the two carbon atoms (marked by the yellow dots) that were the nearest neighbors of the lost carbon atom. The same process annihilates the original 5-6 pair and creates a new 5-6 pair as termination. The structure shown in panel **b** shows no clear dimerization pattern as a result of structural resonance involving several degenerate configurations (more details are in the main text).

**Fig. 4 Electronic, transport and valley-filtering properties of the line defect**

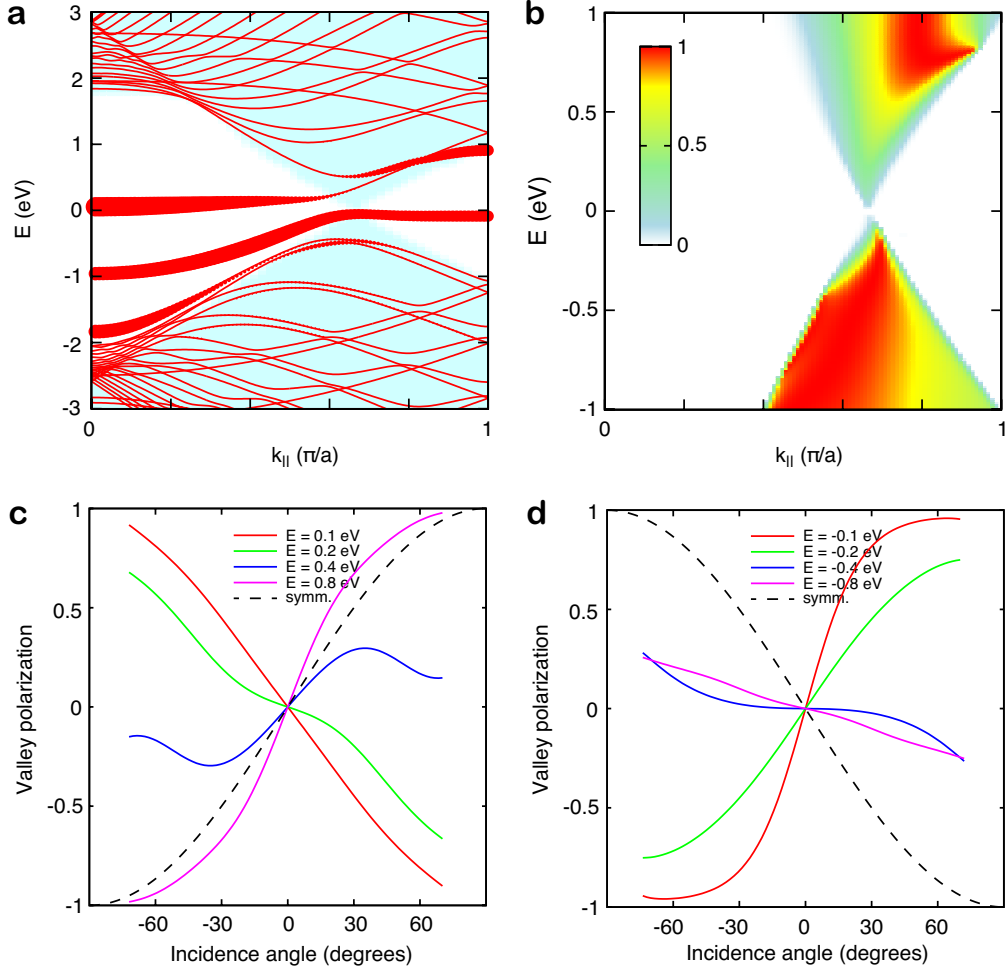

**Fig. 4 | Electronic, transport and valley-filtering properties of the line defect.** **a**. First-principles electronic band structure calculated for the model of line defect along $k_\parallel$ at $k_\perp = 0$. The circles indicate the degree of localization of electronic states at the line defect. The shaded area corresponds to the continuum of bulk graphene states projected onto the 1D Brillouin zone of the line defect. **b**. Transmission probability through the line defect as a function of charge-carrier momentum $k_\parallel$ and energy $E$. **c**,**d**. Valley polarizations calculated for electrons and holes, respectively, as a function of the incident angle of charge carriers at their different energies. The dashed lines correspond to the symmetry-based model of Ref. 7.

**Fig. 5 Proposed electrically operated graphene valley valve**

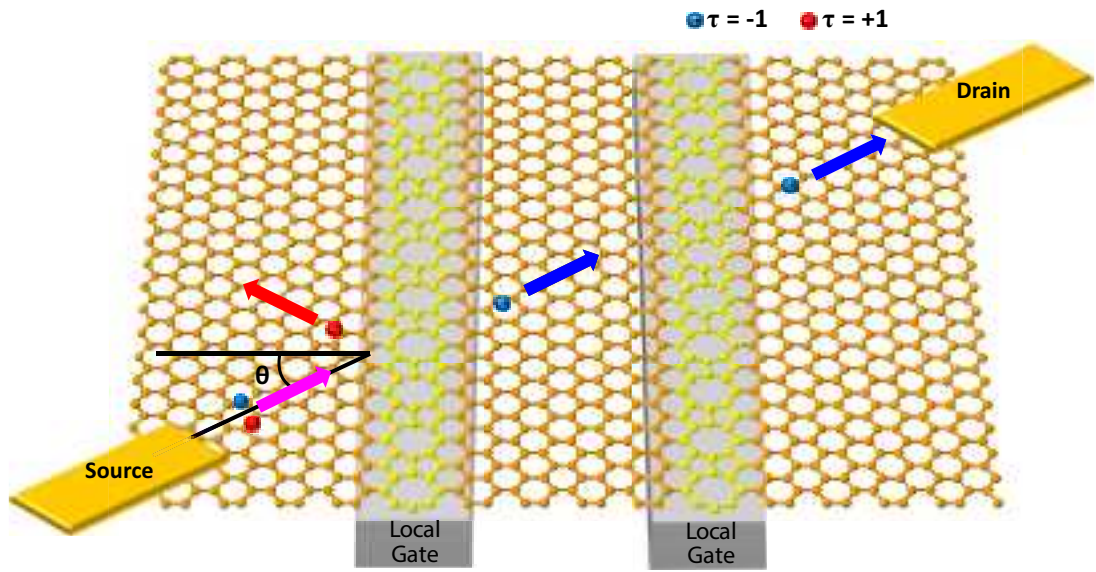

**Fig. 5 | Proposed electrically operated graphene valley valve.** Schematic drawing of an electrically operated graphene valley valve utilizing a pair of parallel 5-5-8 line defects. Electron charge carriers without net valley polarization are injected at a non-zero angle towards two parallel line defects. The Fermi level of the two line-defect regions are independently controlled by two local gates so as to generate and detect valley polarization in the left and right line defects, respectively.

# Supplementary Information for

# Controlled Growth of a Line Defect in Graphene and implications for Gate-Tunable Valley Filtering

J.-H. Chen[1,2,¶], G. Autès[3], N. Alem[1,2,δ], F. Gargiulo[3], A. Gautam[4], M. Linck[4,5], C. Kisielowski[4], O. V. Yazyev[3], S. G. Louie[1,2] and A. Zettl[1,2]

[1]Department of Physics, University of California at Berkeley, Berkeley, CA 94720, USA

[2]Materials Science Division, Lawrence Berkeley National Laboratory, Berkeley, CA 94720, USA

[3]Institute of Theoretical Physics, Ecole Polytechnique Fédérale de Lausanne (EPFL), CH-1015 Lausanne, Switzerland

[4]National Center for Electron Microscopy, Lawrence Berkeley National Laboratory, Berkeley, CA 94720, USA

[5]CEOS-GmbH, Englerstr. 28, D-69126 Heidelberg, Germany

*Correspondence to:  azettl@berkeley.edu

[¶]Current Address: International Center for Quantum Materials, School of Physics, Peking University, Beijing, China 100871

[δ]Current Address:  Department of Materials Science and Engineering, Pennsylvania State University, University Park, PA 16802, USA

[†]Current Address: CEOS-GmbH, Englerstr. 28, D-69126 Heidelberg, Germany

## 1. Temperature Estimations

The temperature of graphene under large current bias during the experiment was estimated from finite-element simulation using Ansys Workbench.  Graphene/Silicon Nitride bilayer of 500 μm × 500 μm was used in the model.  The detail parameters are: the thermal conductivity of graphene $\kappa_{Graphene} = \frac{5000}{1+0.01(T-350K)}$ Wm$^{-1}$K$^{-1}$ [1] and silicon nitride thin film $\kappa_{SiliconNitride} = 3.365$ Wm$^{-1}$K$^{-1}$ (the mean value from Refs. [2] and [3]); the electrical resistivity of the CVD graphene film is measured to be $\rho_{Graphene} = 2.5$ kΩ/☐ and current density $j_{Graphene} \approx 1.5 \times 10^{10}$ A/m$^2$, assuming a thickness of graphene to be 0.36 nm; the thermal resistivity at the graphene-silicon nitride interface $r_{G-S} = 4.2 \times 10^{-8}$ Km$^2$W$^{-1}$ [1], and the emissivity of graphene $\varepsilon_{Graphene} = 0.023$ [1, 4] and silicon nitride $\varepsilon_{SiliconNitride} = 0.2$ [5]. The boundary conditions are: the temperature at the boundary of 500 μm × 500 μm square was set to be 298K, while graphene and silicon nitride radiates into an environment of 298K, convection is set to be zero since the sample was under high vacuum during the experiment. The model geometry and the temperature distribution are shown in Figures S1a and S1b. Note that the thickness of the graphene/silicon nitride bilayer is not to scale. All the data was collected in the central region where the temperature of the graphene sample is about 1300K. Using the resistivity and current density data from above, the total power input to the system is calculated as 18.225 mW. The total heat flux out of the four edges of the graphene-silicon nitride bilayer can be estimated from the simulation to be about 13 mW, indicating that about 5 mW was dissipated through radiation.

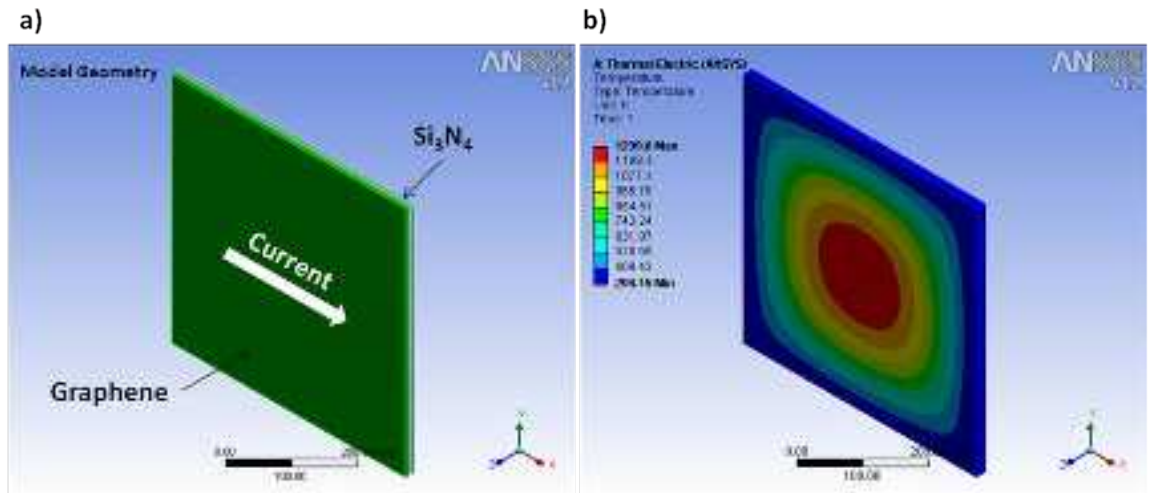

Figure S1. Finite element simulation of the temperature of graphene sample during the electrical biasing experiment. a) The graphene/silicon nitride bilayer model; b) simulated temperature distribution of the graphene device during experimental conditions.

## 2. Instrumentation and imaging conditions

The TEAM 0.5 microscope, which is equipped with Cs corrector, was used in this study. All the images in this publication were taken at 80 kV. To enable atomic resolution at the low accelerating voltage, the resolution-limiting effect of chromatic aberration was reduced by using a monochromator to limit the energy width in the illuminating electron beam. For imaging, a focal series of images were recorded by TEAM 0.5 microscope and the electron exit wave was reconstructed using the MacTempas image processing and simulation software. The phase of the electron exit wave was reconstructed [6] from the focal series using the following parameters:

Third and fifth order spherical aberration: $C_3 = -6$ μm,  $C_5 = 4.5$ mm,

Focus step size = 10 Å.

To further correct for the residual lens aberrations, two fold astigmatism of 120 Å in the direction of 134 degrees, and three fold astigmatism of 700 Å in the direction of 0 degree were introduced into the propagated wave. In addition, coma was set to 183 Å in the direction of 50 degrees.

## 3. Calculation of the formation energies of 7-7-5 and 5-6 termination structures

In order to investigate the formation energies of different defect terminations, we performed calculations on the finite segments of line defect of varying length. Due to the long-ranged strain fields produced by the studied defects, sufficiently large models have to be considered. In particular, in our simulations the defects were embedded in a rectangular periodic supercell of graphene containing 1200 atoms. Models of this size are difficult to treat using first-principles techniques, thus the present simulations have been performed using the classical force field approach implemented in the LAMMPS package [7]. We used the AIREBO potential.

Two structural models of short line defects with 5-6 and 7-7-5 terminations characterized by formation energies of 18.13 eV and 14.97 eV, respectively, are shown in Figs. S2a and S2b. We found that equal-length line defects with 7-7-5 terminations are generally more stable than the 5-6-terminated line defects by ~3 eV (i.e. by ~1.5 eV per termination). Figure S2c shows the dependence of the formation energy of 7-7-5-terminated line defect on its length expressed in terms of the number of 5-5-8 structural units. While the smallest defect considered has formation energy of 14.97 eV, its elongation by each additional 5-5-8 unit increases the formation energy by 4.42 eV.

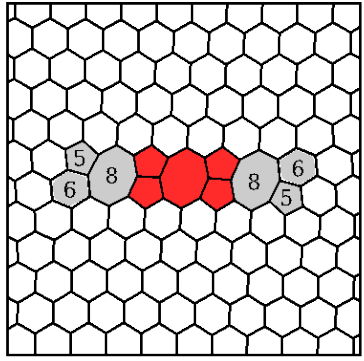 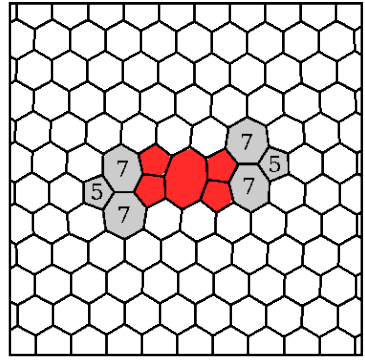

a) 5-6 termination - $E_{form}$=18.13eV  b) 7-7-5 termination - $E_{form}$=14.97eV

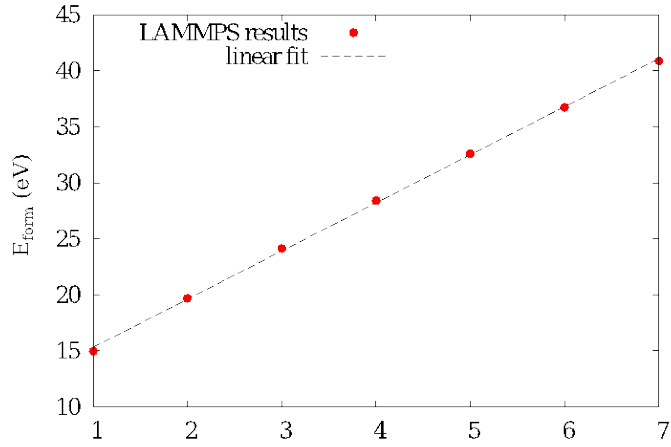

c)

Figure S2. Atomic structures of short (a) 5-6- and (b) 7-7-5-terminated line defects. (c) Formation energy of a 7-7-5-terminated line defect as a function of its length expressed in terms of the number of 5-5-8 structural units.

# References

[1] M. Freitag, M. Steiner, Y. Martin, V. Perebeinos, Z. Chen, J. C. Tsang, and P. Avouris, Nano Lett. 9, 1883 (2009).

[2] P. Eriksson, J. Y. Andersson, and G. Stemme, J. Microelectromechanical Syst. 6, 55 (1997).

[3] M. von Arx, O. Paul, and H. Baltes, J. Microelectromechanical Syst. 9, 136 (2000).

[4] R. R. Nair, P. Blake, A. N. Grigorenko, K. S. Novoselov, T. J. Booth, T. Stauber, N. M. R. Peres, and A. K. Geim, Science 320, 1308 (2008).

[5] F. Völklein, Thin Solid Films 188, 27 (1990).

[6] A. Thust, W. M. J. Coene, M. Op de Beeck, and D. Van Dyck, Ultramicroscopy 64, 211 (1996).

[7] S. Plimpton, J. Comp. Phys. 117, 1 (1995).